\documentclass[twocolumn,aps,pre,floatfix,superscriptaddress,longbibliography]{revtex4-1}
\pdfoutput=1 
\usepackage{amsmath,amssymb,eucal,graphicx}

\usepackage[colorlinks=true, urlcolor=blue, anchorcolor=blue, citecolor=blue,filecolor=blue,linkcolor=blue,menucolor=blue,pagecolor=blue]{hyperref}

\begin{document}
\title{Diffusion in a fluid flow generated by a source at the apex of a wedge}

\author{P. L. Krapivsky}
\affiliation{Department of Physics, Boston University, Boston, MA 02215, USA}
\affiliation{Santa Fe Institute, 1399 Hyde Park Road, Santa Fe, NM 87501, USA}

\begin{abstract}
We consider a particle diffusing inside a wedge with absorbing boundaries and driven by a radial flow of incompressible fluid generated by a source at the apex. The survival probability decays as (time)$^{-\beta}$ with $\beta$ depending on the opening angle of the wedge and the Reynolds number associated with the hydrodynamic flow. The computation of the exponent $\beta$ reduces to finding the ground state energy of the quantum particle in an infinitely deep potential well with a shape determined by the radial flow velocity.
\end{abstract}
\maketitle

\section{Introduction}

The problem of survival of a Brownian particle inside a wedge with absorbing boundaries has been investigated by Sommerfeld \cite{Sommerfeld} in the $19^\text{th}$ century. This and similar problems are encountered in various settings \cite{SR,Comtet03,Lagache,Esguerra08,Esguerra13,OB15}. For instance, suppose we seek the probability that $N$ one-dimensional Brownian particles do not meet during the time interval $(0,t)$. This is equivalent to the probability that a single Brownian particle remains confined to the conical region \hbox{$x_1<x_2<\ldots<x_N$} in $\mathbb{R}^N$. For three particles, the corresponding single Brownian particle remains inside a wedge formed by two intersecting planes (the angle of the wedge depends on the ratios of diffusion coefficients  \cite{SR}). Other properties of Brownian particles often admit a similar interpretation in terms of a single Brownian particle confined to a conical region \cite{F84,HF84,FG88,KR96,KR99,Grab99,Dani-leader,CK03,BK10a,BK10b}, and this conical region is a wedge in the case of three particles. The properties of diffusion inside a wedge with absorbing boundaries also explain the long-time kinetics of  several one-dimensional reaction-diffusion processes \cite{SR,DbA,book}.

In this paper we analyze the behavior of a Brownian particle that is advected by a flow of incompressible fluid generated by the source or sink at the apex of the wedge, and absorbed on the boundaries of the wedge. The velocity field is purely radial. The mathematical description of the processes in unidirectional flow fields is significantly simpler than in the case of generic flows. Therefore the phenomena in unidirectional flow fields have been investigated in numerous contexts such as microfluidics \cite{Kirby,Tabeling}, transport in porous media \cite{Torquato}, stochastic processes \cite{SR,Schuss15}, biological reactions \cite{Linderman}, electrokinetic phenomena \cite{Hunter}, etc. Most analyses consider flows in the pipe \cite{Mortensen05,Bazant16,Chevalier21} that are unidirectional and {\em uniform}, that is, independent on the longitudinal coordinate. The flow field in the wedge is unidirectional (radial) but non-uniform, albeit the $r^{-1}$ variation with the distance from the apex is still sufficiently simple that helps in the analysis.

We consider the wedge with an opening angle $2\alpha$ and absorbing boundaries. The particle eventually gets absorbed, and the probability that it has survived during time interval $(0,t)$ has an algebraic $t^{-\beta}$ long-time tail. In the case of no-flow, $Q=0$, the persistence exponent $\beta=\pi/(4\alpha)$ is well-known, see e.g. \cite{SR,DbA,book}. 

For the ideal incompressible fluid, the velocity field is $v=Q/(2\alpha r)$ where $Q$ is the strength of the source. The exponent $\beta$ in this situation is given by (Sec.~\ref{sec:ideal})
\begin{equation}
\label{beta-ideal}
 \beta=\frac{\sqrt{4\pi^2 D^2+Q^2} - Q}{8\alpha D}
\end{equation}
where $D$ is the diffusion constant.

For the viscous incompressible fluid, the velocity remains radial, $v=u(\theta)/r$. The Navier-Stokes equations are solvable in this situation \cite{Batchelor,LL-FM,Wang91,Drazin}. The exact solution for the viscous flow in the wedge goes back to Jeffery \cite{Jeffery} and Hamel \cite{Hamel}. The Jeffery-Hamel solution is tricky \cite{Harrison,Dean,Rosenhead,LL-FM} in the case of the source ($Q>0$) when its validity is questionable for sufficiently large Reynolds numbers. Computing the decay exponent $\beta$ is mathematically equivalent (Sec.~\ref{sec:viscous}) to finding the ground state energy of the quantum particle in an infinitely deep potential well: $U(\pm \alpha)=\infty$ and  $U(\theta)$ proportional to $u(\theta)$ when $|\theta|<\alpha$. The analogy with Schr\"{o}dinger equation describing stationary states of a quantum particle in a one-dimensional potential is amusing, albeit we do not use it in computations. What is needed are the basic properties of the Jeffery-Hamel solution describing the viscous flow in the wedge. These properties are outlined in Sec.~\ref{sec:JH}. In Secs.~\ref{sec:convergent}--\ref{sec:small} we employ perturbation techniques and deduce analytical predictions for $\beta$ in the limiting cases of high and low Reynolds numbers. Three-dimensional analogs of the wedge problem, particularly the diffusion-advection in the jet flow, are briefly discussed in Sec.~\ref{sec:3d}.  Conclusions are presented in Sect.~\ref{sec:concl}.

\section{Ideal Incompressible Flows in Wedges}
\label{sec:ideal}

At time $t=0$, the particle is released inside the wedge. In polar coordinates $(r,\theta)$, the wedge is the region $r\geq 0$ and $|\theta|\leq \alpha$ with $2\alpha$ being the opening angle of the wedge. Let $S(r,\theta,t)$ be the probability that the particle released from $(r,\theta)$ has not touched the boundaries of the wedge during the time interval $(0,t)$.  This survival probability satisfies
\begin{equation}
\label{S-eq}
\partial_t S=D\nabla^2 S + ({\bf v}\cdot\nabla)S
\end{equation}
Here $D$ is the diffusion constant and ${\bf v}$ the velocity field. The advection term $({\bf v}\cdot\nabla)S$ is on the right-hand side  because $S(r,\theta,t)$ depends on the {\em initial} position of the particle, and therefore Eq.~\eqref{S-eq} is the {\em backward} advection-diffusion equation.

The velocity field of ideal incompressible fluid generated by a source of strength $Q$ reads
\begin{equation}
\label{v2d}
  \mathbf{v}(\mathbf{r})=\frac{Q}{2\alpha r}\,\mathbf{\hat r}
\end{equation}
In rotationally-symmetric situations with velocity field inversely proportional to the distance from the origin, advection can be absorbed into diffusion by an appropriate shift of the spatial dimension \cite{Bray00,KR07,PK12,KR18}. The advection-diffusion equation \eqref{S-eq} with velocity field \eqref{v2d} becomes
\begin{equation}
\label{S-ideal}
\frac{\partial S}{\partial t}=
D\left(\frac{\partial^2 S}{\partial r^2} + \frac{1+\text{Pe}}{r}\,\frac{\partial S}{\partial r} + \frac{1}{r^2}\,\frac{\partial^2 S}{\partial \theta^2}\right)
\end{equation}
where $\text{Pe}=\frac{Q}{2\alpha D}$ is the P\'{e}clet number. In a rotationally-symmetric situation in $d$ dimensions, the diffusion equation reads
\begin{equation}
\label{Sd}
\frac{\partial S}{\partial t}=
D\left(\frac{\partial^2 S}{\partial r^2} + \frac{d-1}{r}\,\frac{\partial S}{\partial r} + \frac{1}{r^2}\,\frac{\partial^2 S}{\partial \theta^2}\right)
\end{equation}
Equations \eqref{S-ideal} and \eqref{Sd} coincide if we identify $d=2+\text{Pe}$.  Thus the two-dimensional advection-diffusion process with a point sink or source of mass may be recast as a pure diffusion process in a fictitious space of dimension $d=2+\text{Pe}$.

An exact analysis of the linear governing equation \eqref{Sd} subject to the initial condition $S(r,\theta, 0)=1$ and the absorbing boundary conditions $S(r, \pm \alpha,t)=0$ is possible. Our chief goal, however, is the extracting of the large time behavior, and we shall take advantage of the useful feature of the survival probability in this limit, namely its algebraic decay:
\begin{equation}
\label{decay}
S(r,\theta,t) \simeq \,\Phi(r,\theta)\, t^{-\beta}
\end{equation}
Substituting \eqref{decay} into the advection-diffusion equation \eqref{S-ideal} and noting that the time derivative becomes negligible in the long-time limit we find that $\Phi(r,\theta)$ satisfies
\begin{equation}
\label{Phi}
\frac{\partial^2 \Phi}{\partial r^2} + \frac{1+\text{Pe}}{r}\,\frac{\partial \Phi}{\partial r} + \frac{1}{r^2}\,\frac{\partial^2 \Phi}{\partial \theta^2} = 0
\end{equation}
The dependence of $\Phi(r,\theta)$ on the radial coordinate can be determined using dimensional analysis. In principle,  $\Phi(r,\theta)=\Phi(r,\theta | D, \text{Pe})$, so $\Phi$ depends on two dimension-full quantities $r$ and $D$. The dimension of $\Phi$ is $T^\beta$ where $T$ denotes the dimension of time; this is obvious from \eqref{decay} since the survival probability is dimensionless. The only variable with dimension $T^\beta$ which can be composed of $r$ and $D$ is $(r^2/D)^\beta$.  Thus
\begin{equation}
\label{Phi-sep}
\Phi(r,\theta)=\left(\frac{r^2}{D}\right)^{\beta}\psi(\theta)
\end{equation}
Plugging \eqref{Phi-sep} into \eqref{Phi} we obtain
\begin{equation}
\label{phi}
\psi''(\theta)+ \big[4\beta^2+2\beta\, \text{Pe}\big]\psi(\theta) = 0
\end{equation}
where prime denotes the derivative with respect to $\theta$. There are infinitely many linearly-independent solutions of \eqref{phi} satisfying the boundary conditions
\begin{equation}
\label{BC}
\psi(\pm\alpha)=0
\end{equation}
describing absorbing boundaries of the wedge. These solutions are $\psi_n=\cos(\lambda_n \theta)$ with $\lambda_n=\pi(2n+1)/(2\alpha)$ and $n=0,1,2,\ldots$. The physical requirement of positivity, $\psi(\theta)>0$  when $|\theta|<\alpha$, allows us to select the unique physically relevant solution
\begin{equation}
\label{GS}
\psi=\cos\!\left(\frac{\pi \theta}{2\alpha}\right)
\end{equation}
and determine the decay exponent
\begin{equation}
\label{beta:a-q}
\beta = \frac{\sqrt{(\pi/\alpha)^2 + \text{Pe}^2} - \text{Pe}}{4}
\end{equation}
This formula coincides with the announced result \eqref{beta-ideal}. From \eqref{beta-ideal} or \eqref{beta:a-q} we recover the well-known  \cite{SR} expression $\beta=\frac{\pi}{4\alpha}$ in the no-flow case ($\text{Pe}=0$).

\begin{figure}[ht]
\begin{center}
\includegraphics[width=0.4\textwidth]{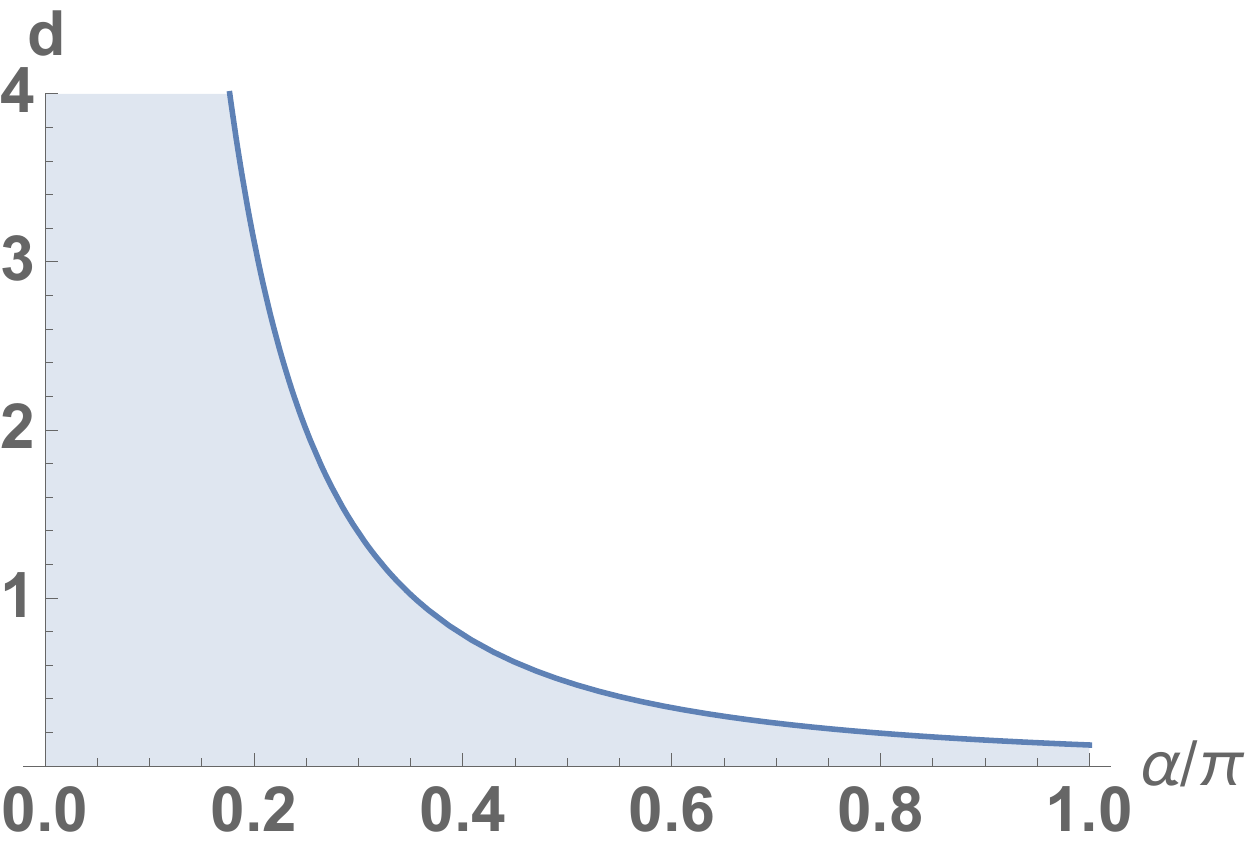}
\caption{The average lifetime of the Brownian particle is finite in the filled region defined by \eqref{d-alpha}.}
\label{fig:d-alpha}
  \end{center}
\end{figure}

The mean exit time of the Brownian particle
\begin{equation}
T(r,\theta) = \int_0^\infty dt\,t\left[-\frac{dS}{dt}\right]= \int_0^\infty dt\,S(r,\theta,t)
\end{equation}
diverges when $\beta\leq 1$; for $\beta>1$, the mean exit time is finite. The boundary between these two regimes is determined from $\beta=1$ which in conjunction with \eqref{beta:a-q} yield $\text{Pe}+2=\pi^2/(8\alpha^2)$. Recalling that the effective dimension of the pure diffusion process with the same decay exponent is $d=\text{Pe}+2$ we conclude that the mean exit time is finite when (see also Fig.~\ref{fig:d-alpha})
\begin{equation}
\label{d-alpha}
d < \frac{\pi^2}{8\alpha^2}
\end{equation}

If $\beta>1$, the mean exit time $T(r,\theta)$ satisfies
\begin{equation}
\label{Trt}
\frac{\partial^2 T}{\partial r^2} + \frac{1+\text{Pe}}{r}\,\frac{\partial T}{\partial r} + \frac{1}{r^2}\,\frac{\partial^2 T}{\partial \theta^2} = -\frac{1}{D}
\end{equation}
(see e.g. \cite{KR18}). The dependence of the mean exit time on the radial coordinate is fixed by dimensional analysis
\begin{equation}
\label{T-sep}
T(r,\theta)=\frac{r^2}{D}\,\tau(\theta)
\end{equation}
Plugging \eqref{T-sep} into \eqref{Trt} we obtain
\begin{equation}
\label{tau}
\tau''+ (4 + 2\text{Pe})\tau = -1
\end{equation}
Solving \eqref{tau} subject to $\tau(\pm \alpha)=0$ and selecting the physically relevant solution satisfying $\tau(\theta)>0$  when $|\theta|<\alpha$ we arrive at the neat expression for the mean exit time
\begin{equation}
\label{T:ideal}
T(r,\theta)=\frac{r^2}{D}\, \frac{\cos(p\theta)-\cos\big(p\alpha)}{p^2\,\cos(p\alpha)}\,, \quad p\equiv \sqrt{4 + 2\text{Pe}}
\end{equation}
applicable when $p\alpha<\frac{\pi}{2}$, equivalently $\text{Pe}+2<\pi^2/(8\alpha^2)$, which is exactly the requirement that $\beta>1$.

First passage characteristics of a Brownian particle advected by ideal incompressible flows have been investigated in several studies, particularly in two dimensions where one can use  complex analysis, conformal mappings, the Wiener-Hopf technique, etc. (see, e.g., \cite{Krook,Koplik,Bazant04}).  First passage characteristics of a Brownian particle advected by viscous incompressible flows can be investigated analytically in a few situations since exact solutions of the Navier-Stokes equations are rare \cite{Batchelor,LL-FM,Wang91,Drazin}. Most of such exact solutions are unidirectional flows \cite{Wang91,Drazin,Bazant16}. The viscous flow in the wedge is one such solvable case with purely radial flow.

\section{Viscous Incompressible Flows in Wedges: First Passage Characteristics}
\label{sec:viscous}

In the case of incompressible viscous fluid, the velocity field generated by the source at the apex of the wedge is radial and inversely proportional to the distance from the apex. The chief distinction from the ideal flow is that the velocity now depends on $\theta$. The incompressible viscous flow in the wedge was first studied by Jeffery \cite{Jeffery} and Hamel \cite{Hamel}. This flow represents a rare exact solution of the Navier-Stokes equations of the incompressible viscous fluid. The solvability of the Navier-Stokes equations is the consequence of the uni-direction nature of the flow and simple dependence on the radial coordinate:
\begin{equation}
\label{Hamel}
v(r,\theta) = \frac{\nu}{r}\,u(\theta)
\end{equation}
The function $u(\theta)$ is dimensionless since we have used the kinematic viscosity $\nu$ as a pre-factor. With velocity field \eqref{Hamel}, the advection-diffusion equation \eqref{S-eq} becomes
\begin{equation}
\label{S-simple}
\frac{\partial S}{\partial t}=
D\left(\frac{\partial^2 S}{\partial r^2} + \frac{1+\sigma u(\theta)}{r}\,\frac{\partial S}{\partial r} + \frac{1}{r^2}\,\frac{\partial^2 S}{\partial \theta^2}\right)
\end{equation}
where $\sigma = \nu/D$ is the Prandtl number.

Using the ansatz \eqref{decay} we find that $\Phi(r,\theta)$ satisfies
\begin{equation}
\label{Phi-eq}
\frac{\partial^2 \Phi}{\partial r^2} + \frac{1+\sigma u(\theta)}{r}\,\frac{\partial \Phi}{\partial r} + \frac{1}{r^2}\,\frac{\partial^2 \Phi}{\partial \theta^2} = 0
\end{equation}
Seeking again the solution in the form \eqref{Phi-sep} we arrive at
\begin{equation}
\label{phi-eq}
\psi''(\theta)+ \big[4\beta^2+2\beta\sigma u(\theta)\big]\psi(\theta) = 0
\end{equation}

One can think about this linear ordinary differential equation as a Schr\"{o}dinger equation describing stationary states of a particle in a one-dimensional potential \cite{LL-QM}. Finding the exponent $\beta$ constitutes a (non-linear) eigenvalue problem. The solution must be positive inside the wedge, so we are seeking the ground state of a quantum particle in an infinitely deep potential well $-\alpha<\theta<\alpha$. The potential is
\begin{equation}
U(\theta) =
\begin{cases}
-\sqrt{E}\,\sigma u(\theta)   & |\theta| < \alpha \\
\infty                                   & |\theta| = \alpha
\end{cases}
\end{equation}
where $E=4\beta^2$ is the energy of the ground state. Inside the well, the potential is $-\sqrt{E}\,\sigma u(\theta)$. The function $u(\theta)$ is expressible in terms of elliptic integrals \cite{Jeffery,Hamel}. The potential is also proportional to $\sqrt{E}$ and it then determines the energy $E$ of the ground state, so we effectively have a non-linear eigenvalue problem. Finding a ground state is analytically impossible due to the complicated nature of $u(\theta)$. In the limiting situations of high and low Reynolds numbers, one can employ asymptotic methods to obtain perturbative results as we show in Secs.~\ref{sec:convergent}--\ref{sec:small}.

If $\beta>1$, the mean exit time $T(r,\theta)$ satisfies
\begin{equation}
\label{Tr-viscous}
\frac{\partial^2 T}{\partial r^2} + \frac{1+\sigma u(\theta)}{r}\,\frac{\partial T}{\partial r} + \frac{1}{r^2}\,\frac{\partial^2 T}{\partial \theta^2} = -\frac{1}{D}
\end{equation}
The ansatz \eqref{T-sep} remains applicable, so  \eqref{Tr-viscous} simplifies to
\begin{equation}
\label{tau-viscous}
\tau''(\theta)+ [4+2\sigma u(\theta)]\tau(\theta) = -1
\end{equation}

\section{Jeffery-Hamel Solution}
\label{sec:JH}

Here we outline a few properties of the Jeffery-Hamel viscous flow necessary for the analysis of advection-diffusion in the wedge. The velocity field \eqref{Hamel} has a single radial component, so it manifestly satisfies the continuity equation for the incompressible fluid.  Navier-Stokes equations \cite{Batchelor,LL-FM} reduce to
\begin{subequations}
\begin{align}
\label{vr}
&v\,\frac{\partial v}{\partial r}+\frac{\partial p}{\partial r}=
\nu\left(\frac{\partial^2 v}{\partial r^2} + \frac{1}{r}\,\frac{\partial v}{\partial r} -\frac{v}{r^2}+ \frac{1}{r^2}\,\frac{\partial^2 v}{\partial \theta^2} \right)\\
\label{pv}
&\frac{\partial p}{\partial \theta}=\frac{2\nu}{r}\,\frac{\partial v}{\partial \theta}
\end{align}
\end{subequations}
for purely radial two-dimensional flows. (The density of an incompressible fluid is constant, we have set it to unity without loss of generality.)

Using \eqref{Hamel}, one re-writes \eqref{pv} as
\begin{equation*}
\frac{\partial p}{\partial \theta}=\frac{2\nu^2}{r^2}\,\frac{d u}{d \theta}
\end{equation*}
which is integrated to yield
\begin{equation}
\label{pf}
p = \frac{2\nu^2}{r^2}\,u(\theta) + f(r)
\end{equation}
Substituting \eqref{Hamel} and \eqref{pf} into \eqref{vr} we obtain
\begin{equation}
\label{fu}
r^3\,\frac{df}{dr} = \nu^2\left[u''+4u+u^2\right]
\end{equation}
The left-hand side of \eqref{fu} depends on $r$, while the right-hand side depends on $\theta$. Hence both sides must be equal  to the same constant. Multiplying
\begin{equation}
\label{u-eq}
u''+4u+u^2= C_1
\end{equation}
by $2u'$ and integrating one obtains
\begin{subequations}
\begin{equation}
\label{u1}
(u')^2+4u^2+\frac{2u^3}{3} = 2C_1 u + C_2
\end{equation}
The boundary conditions are
\begin{equation}
\label{u:BC}
u(\pm\alpha)=0
\end{equation}
\end{subequations}
The solution of the boundary-value problem \eqref{u1}--\eqref{u:BC} can be written in terms of elliptic integrals.  These results were already known to Jeffery \cite{Jeffery} and Hamel \cite{Hamel}. The problem is more rich \cite{Harrison,Dean,Rosenhead,LL-FM} than it was initially believed. For instance, depending on $\alpha$, the Reynolds number $\text{Re}=|Q|/\nu$ and the sign of $Q$ there can be situations with infinitely many solutions; see \cite{Fraenkel62,Banks88} for the classification of solutions.

Simpler behaviors occur if the volume flux
\begin{equation}
\label{Q-u}
Q = \nu\int_{-\alpha}^\alpha d\theta\,u(\theta)
\end{equation}
is negative, that is, the flow is generated by a sink; flows in a wedge generated by sinks are also known as flows in converging channels.
A convergent symmetric flow, that is a solution with everywhere negative $u(\theta)$ satisfying $u(\theta)=u(-\theta)$, exists for any $Q<0$ if $\alpha<\frac{\pi}{2}$. This has been established in \cite{Rosenhead,LL-FM} for convergent flows with arbitrary Reynolds numbers. For small  Reynolds numbers, $|Q|/\nu\ll 1$, symmetric solutions with everywhere negative (when $Q<0$) or positive (when $Q>0$) velocity exist for all $\alpha<\alpha_*= 2.2467\ldots$, see Sec.~\ref{sec:small} for details.

If $Q>0$, the symmetric solutions with everywhere positive velocity do not exist when the Reynolds number $\text{Re}=Q/\nu$ is sufficiently large, although there may be symmetric solutions involving regions of inward and outward flow \cite{Rosenhead,Fraenkel62,Banks88}. The stability of some of these solutions has been studied, see \cite{Sobey86,Scott94,Dennis97}. In the physically interesting case of large Reynolds numbers,  $\text{Re}=|Q|/\nu\gg 1$, the main results can be summarized as follows:
\begin{enumerate}
\item Convergent flows approach to the solution of the Euler equations,  \eqref{v2d}, except narrow boundary-layer regions, $\alpha-|\theta|\sim  \text{Re}^{-1/2}$.
\item For divergent flows, the number of alternating minima and maxima diverge as $\text{Re}\to\infty$, so there is no definite limiting solution. In experiments, non-stationary and turbulent flows are observed.
\end{enumerate}

Let us now analyze diffusion in convergent channels at large Reynolds numbers.

\section{Convergent flows at high Reynolds numbers}
\label{sec:convergent}

For sink flows at high Reynolds numbers ($Q<0$ and $\text{Re}\gg 1$), the velocity field is close to the potential flow \eqref{v2d}, so the exact expression \eqref{beta:a-q} for the decay exponent in the ideal (non-viscous) fluid may provide a good approximation. Re-writing \eqref{beta:a-q} as
\begin{equation}
\beta_\text{ideal} = \frac{\sigma \text{Re}}{8\alpha} +\sqrt{\left(\frac{\sigma \text{Re}}{8\alpha}\right)^2 + \left(\frac{\pi}{2\alpha}\right)^2}
\end{equation}
one deduces
\begin{equation}
\label{b-Re}
\beta_\text{ideal} = \sigma\, \frac{\text{Re}}{4\alpha}  +\frac{\pi^2}{4\sigma  \alpha}\,(\text{Re})^{-1}+O(\text{Re}^{-3})
\end{equation}
when $\text{Re}\gg 1$. The leading term in \eqref{b-Re} is asymptotically exact, but the sub-leading correction decaying as $(\text{Re})^{-1}$ is erroneous. The asymptotically exact sub-leading correction is an increasing function of the Reynolds number that scales as $(\text{Re})^{1/2}$ as we show below. More precisely
\begin{equation}
\label{b-Re-exact}
\beta = \sigma\, \frac{\text{Re}}{4\alpha} +  \sigma\, \frac{3-\sqrt{6}}{\alpha} \sqrt{\frac{\text{Re}}{4\alpha}}+\ldots
\end{equation}

We derive \eqref{b-Re-exact} using perturbation techniques \cite{BO}. We need an expression for $u(\theta)$ when $\text{Re}\gg 1$. An exact solution of the boundary-value problem \eqref{u1}--\eqref{u:BC} simplifies in the $\text{Re}\to \infty$ limit: 
\begin{eqnarray}
\label{uniform}
\frac{u(\theta)}{2B^2} &=& 2 + 3 \tanh^2[2B\alpha+\phi] - 3\tanh^2[B(\alpha-\theta)+\phi]\nonumber\\
                                   &-& 3\tanh^2[B(\alpha+\theta)+\phi]
\end{eqnarray}
where
\begin{equation}
\phi\equiv \tanh^{-1}\sqrt{\frac{2}{3}} = 1.14621583\ldots
\end{equation}
The expression \eqref{uniform} for the flow field is uniformly valid in the entire $|\theta| < \alpha$ region and it reveals the presence of
the boundary layers  near the walls. The thickness of these boundary layers is of the order of $B^{-1}$. Using \eqref{uniform} we compute
$Q=\nu\int_{-\alpha}^\alpha d\theta\,u(\theta)$ to yield
\begin{equation}
\label{Re-B}
\text{Re}=-\frac{Q}{\nu}= 4\alpha B^2 -4\left(3-\sqrt{6}\right)B+\ldots
\end{equation}
from which
\begin{equation}
\label{B-Re}
B = \sqrt{\frac{\text{Re}}{4\alpha}} + \frac{3-\sqrt{6}}{2\alpha}+\ldots
\end{equation}

Due to the symmetry, $u(\theta)=u(-\theta)$, it suffices to consider the half wedge: $0\leq \theta\leq \alpha$. In the outer region
\begin{subequations}
\begin{equation}
\label{outer}
B^{-1}\ll \alpha - \theta\leq \alpha
\end{equation}
the function $u(\theta)$ given by \eqref{uniform} is independent on $\theta$ in the leading order:
\begin{equation}
\label{u0}
 u = - 2B^2
\end{equation}
Indeed, the arguments of two hyperbolic tangents in \eqref{uniform} diverge as $\text{Re}\to \infty$. The argument of the third hyperbolic tangent is also large, as $B(\alpha - \theta)\gg 1$ in the outer layer. Thus we obtain \eqref{u0} in the leading order.  A more precise approximation of \eqref{uniform} is 
\begin{equation}
\label{u1}
\frac{u(\theta)}{2B^2} = -1 +12 e^{-\Theta}, \quad \Theta = B(\alpha-\theta)+\phi
\end{equation}
\end{subequations}
showing that the deviation from $u = - 2B^2$ is exponentially small as $\Theta\gg 1$ in the outer layer. 

Inside the boundary layer
\begin{subequations}
\begin{equation}
\label{inner}
0\leq \alpha - \theta\ll 1
\end{equation}
the velocity varies according to
\begin{equation}
\label{BL}
\frac{u}{2B^2} = 2 - 3\tanh^2\Theta
\end{equation}
\end{subequations}

In the outer region \eqref{outer}, the governing Eq.~\eqref{phi-eq} simplifies to $\psi''+ \big[4\beta^2-4\beta\sigma B^2\big]\psi = 0$, from which
\begin{equation}
\label{outer:psi}
\psi = \cos(2b B\theta), \qquad b \equiv \sqrt{(\beta/B)^2-\beta\sigma}
\end{equation}
In the inner region \eqref{inner}, the governing Eq.~\eqref{phi-eq} becomes
\begin{subequations}
\begin{equation}
\label{inner:psi-eq}
\frac{d^2\Psi}{d \Theta^2} + 4\left[b^2+\frac{3\beta\sigma}{\cosh^2\Theta}\right]\Psi = 0
\end{equation}
where $\Psi(\Theta)\equiv \psi(\theta)$. The boundary condition $\psi|_{\theta=\alpha}=0$ yields
\begin{equation}
\label{inner:BC}
\Psi(\Theta=\phi)=0
\end{equation}
\end{subequations}
Equation \eqref{b-Re} implies that the term in the square brackets in Eq.~\eqref{inner:psi-eq} scales as $\text{Re}$ when the Reynolds number is large. This suggests to apply WKB techniques \cite{BO}. The WKB solution to \eqref{inner:psi-eq}--\eqref{inner:BC} reads
\begin{equation}
\Psi \sim \sin\!\Bigg[2\int_\phi^\Theta dx\,\sqrt{b^2+\frac{3\beta\sigma}{\cosh^2\!x}} \Bigg]
\end{equation}
This solution can be re-written as
\begin{subequations}
\begin{equation}
\label{Psi:WKB}
\Psi \sim \sin[2bB(\alpha-\theta)+2\theta_0]
\end{equation}
with
\begin{equation}
\label{theta:WKB}
\theta_0= \int_\phi^\infty dx\bigg[\sqrt{b^2+\frac{3\beta\sigma}{\cosh^2\!x}}- b\bigg]
\end{equation}
\end{subequations}
In \eqref{theta:WKB} we used again the shorthand notation $b$ defined in Eq.~\eqref{outer:psi} and we also replaced the upper limit $\Theta$ by $\infty$ since the matching the inner solution \eqref{Psi:WKB} with the outer solution \eqref{outer:psi} is made when $\Theta\gg 1$.

Massaging the integral in \eqref{theta:WKB} we obtain
\begin{eqnarray*}
\theta_0 &=&  3\beta\sigma\int_\phi^\infty \frac{dx}{\cosh^2\!x}\,\frac{1}{\sqrt{b^2+\frac{3\beta\sigma}{\cosh^2\!x}} + b} \\
              &=&  3\beta\sigma\int_{\sqrt{2/3}}^1 \,\frac{dy}{\sqrt{b^2+3\beta\sigma(1-y^2)} + b}
\end{eqnarray*}
where we made the transformation $x\to y=\tanh x$ and used $\tanh\phi=\sqrt{2/3}$. We do not display an exact cumbersome expression for the last integral and only show the asymptotically exact result. To derive it we notice that $b = O(1)$ in the $\text{Re}\to \infty$ limit as we will confirm a posteriori. Hence we neglect $b$ and arrive at
\begin{equation}
\label{integral}
\theta_0  = C \sqrt{3\beta\sigma}
\end{equation}
with
\begin{equation*}
C  = \int_{\sqrt{2/3}}^1 \,\frac{dy}{\sqrt{1-y^2}} = \frac{\pi}{2}-\arcsin\sqrt{\frac{2}{3}} = 0.6154797\ldots
\end{equation*}

Matching the inner solution \eqref{Psi:WKB} with the outer solution \eqref{outer:psi} yields
\begin{equation}
bB+\theta_0 = \frac{\pi}{4}
\end{equation}
leading to $\sqrt{\beta^2-\beta\sigma B^2}+C\sqrt{3\beta\sigma} = \pi/4$, or equivalently
\begin{equation}
\label{bBC}
\beta = \sigma (B^2+3C^2)  -\frac{\pi}{2}\,C\sqrt{\frac{3\sigma}{\beta}} + \ldots
\end{equation}
Comparing \eqref{bBC} and \eqref{B-Re} which we re-write as 
\begin{equation*}
B^2 = \frac{\text{Re}}{4\alpha} + \frac{3-\sqrt{6}}{\alpha} \sqrt{\frac{\text{Re}}{4\alpha}}+\ldots
\end{equation*}
we conclude that $\beta \simeq \sigma B^2$. This yields the announced result \eqref{b-Re-exact}.

Finally we show how to determine $b$ using \eqref{bBC}:
\begin{equation*}
b^2  = \beta\left[\frac{\beta}{B^2}-\sigma\right]  \simeq   \beta\sigma\left[\frac{B^2+3C^2}{B^2}-1\right]  \simeq 3 C^2 \sigma^2
\end{equation*}
Thus indeed $b = O(1)$ in the $\text{Re}\to \infty$ limit.

\section{Advection-Diffusion at Low Reynolds numbers}
\label{sec:small}

\begin{figure}[ht]
\begin{center}
\includegraphics[width=0.4\textwidth]{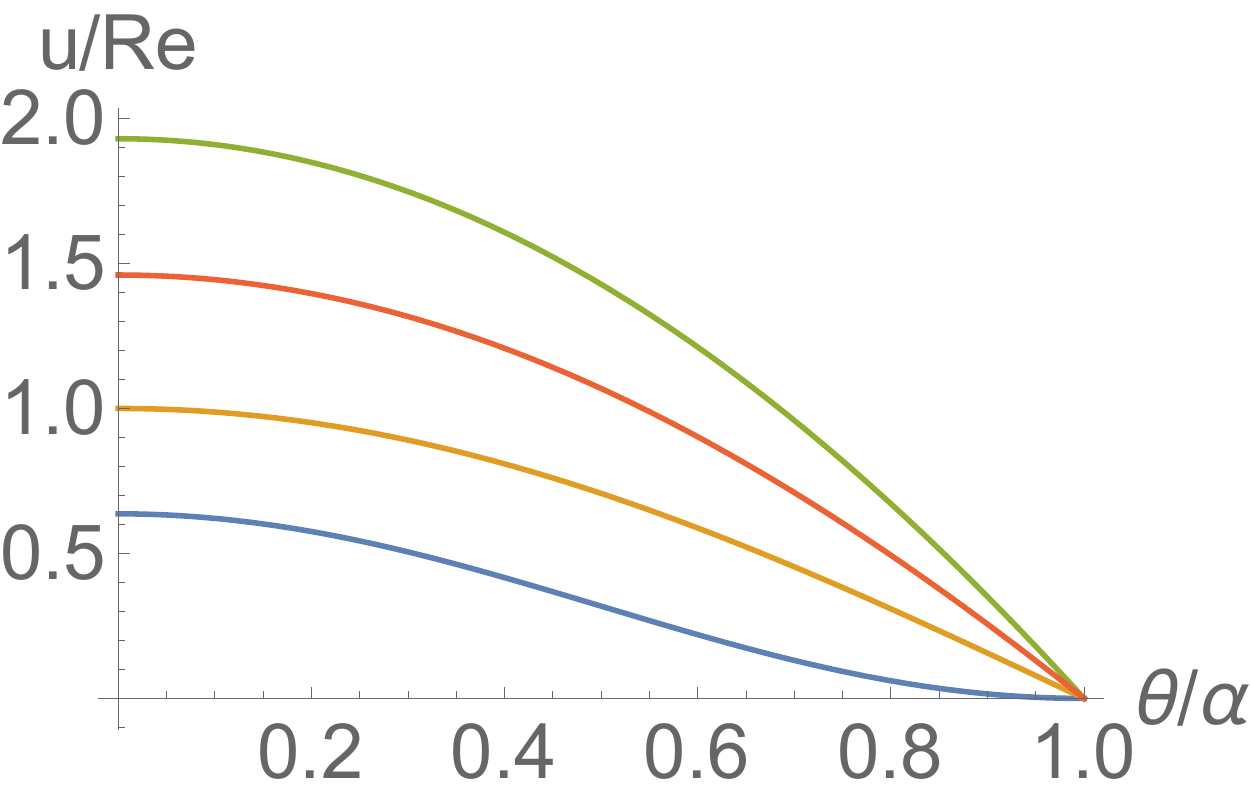}
\caption{The re-scaled velocity $u/\text{Re}$ vs. re-scaled angular coordinate $\theta/\alpha$ for the opening angle
$\alpha=\frac{\pi}{8}, \frac{\pi}{6},\frac{\pi}{4}, \frac{\pi}{2}$ (top to bottom).}
\label{fig:U-plus}
  \end{center}
\end{figure}

\begin{figure}[ht]
\begin{center}
\includegraphics[width=0.4\textwidth]{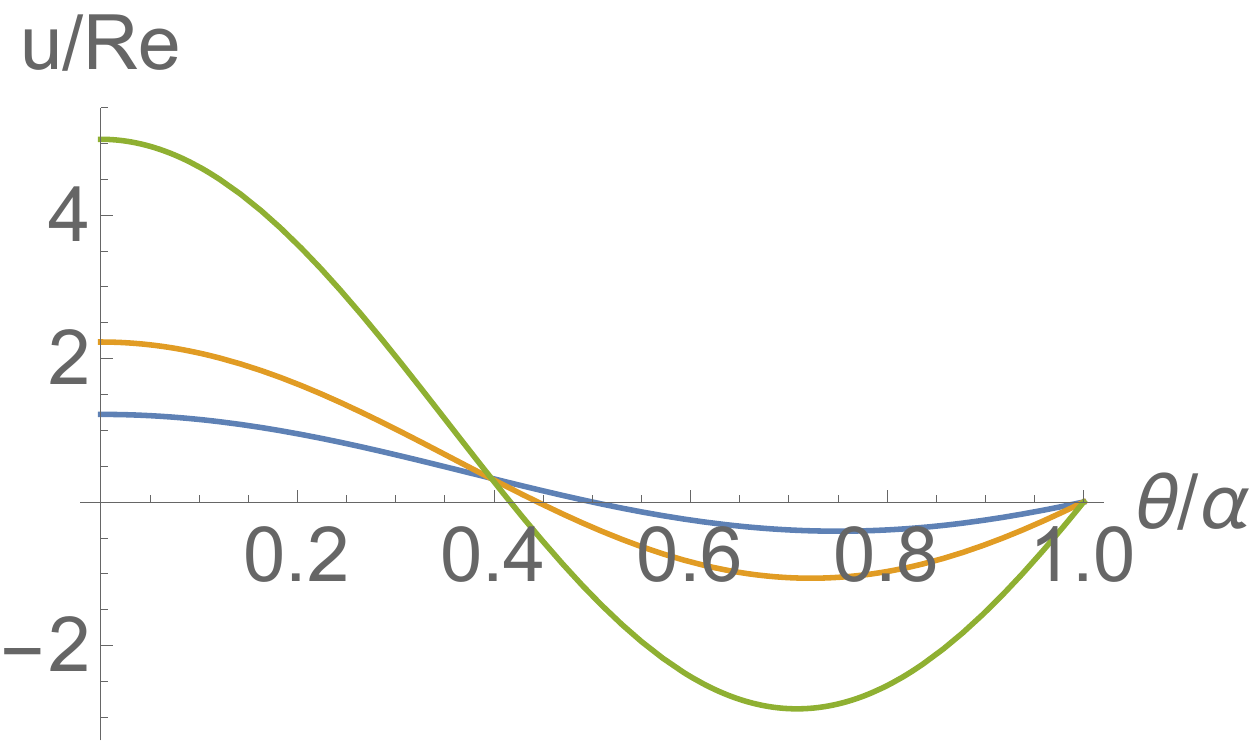}
\caption{The re-scaled velocity $u/\text{Re}$ vs. re-scaled angular coordinate $\theta/\alpha$ for the opening angle
$\alpha=\frac{2\pi}{3}, \frac{9\pi}{13},\frac{12\pi}{17}$. The velocity vanishes inside the wedge, namely at $\theta=\pi-\alpha$.}
\label{fig:U-plus-minus}
  \end{center}
\end{figure}

In the low Reynolds number limit the inertial terms in the Navier-Stokes equations are omitted. In the present situation, we drop the term $v\partial_r v$ on the left-hand side of \eqref{vr}. Instead of \eqref{u-eq} we obtain
\begin{equation}
\label{u-eq-Stokes}
u''+4u= \text{const}
\end{equation}
A solution of Eq.~\eqref{u-eq-Stokes} satisfying the boundary conditions \eqref{u:BC} reads
\begin{equation}
\label{u-Stokes}
u =  \frac{Q}{\nu}\,\frac{\cos(2\theta)- \cos(2\alpha)}{\sin(2\alpha) - 2\alpha\cos(2\alpha)}
\end{equation}
where the amplitude was fixed by \eqref{Q-u}.
The velocity has the same sign as $Q$ inside the wedge when $\alpha\leq \frac{\pi}{2}$, see Fig.~\ref{fig:U-plus}. This is no longer true when $\frac{\pi}{2}<\alpha < \alpha_*$, with $\alpha_*$ defined by by Eq.~\eqref{alpha-star}, see Fig.~\ref{fig:U-plus-minus}.

The denominator in Eq.~\eqref{u-Stokes} vanishes at $\alpha=\alpha_*$ which is the smallest positive root of
\begin{equation}
\label{alpha-star}
\tan(2\alpha_*)=2\alpha_*
\end{equation}
This root is $\alpha_* \approx  2.2467047$, or $\alpha_* \approx 128.727^\circ$. The solution \eqref{u-Stokes} explodes at $\alpha=\alpha_*$, and it is well-defined only when $\alpha<\alpha_*$. Thus the Jeffery-Hamel solutions have rich structure even in the low  Reynolds number limit.

\subsection{Wedge with  $\alpha=\frac{\pi}{4}$}

As a concrete example, consider the wedge with right opening angle, $2\alpha=\frac{\pi}{2}$. The dependence of the velocity on $\theta$ is particularly simple in this case:
\begin{equation}
\label{Stokes}
u = \frac{Q}{\nu}\,\cos(2\theta)
\end{equation}
The governing equation \eqref{phi-eq} turns into an ordinary differential equation
\begin{equation}
\label{Mathieu}
\psi''(\theta)+ \big[4\beta^2+\beta \epsilon \cos(2\theta)\big]\psi(\theta) = 0
\end{equation}
known as the Mathieu equation. Here we shortly write
\begin{equation}
\label{eps}
\epsilon = 2\sigma\, \frac{Q}{\nu} = \frac{2Q}{D}
\end{equation}

We seek a symmetric solution, $\psi(\theta) = \psi(-\theta)$, of Eq.~\eqref{Mathieu} which is positive for all $|\theta|<\frac{\pi}{4}$ and vanishes on the boundaries: $\psi(\pm \pi/4)=0$. In principle, one can solve the problem analytically by using Mathieu functions which arise in numerous problems, see e.g. \cite{MF,QM12}. However, the velocity field \eqref{Stokes} is applicable only at small Reynolds numbers, i.e., $|\epsilon|\ll 1$, and it suffices to determine a perturbative solution of \eqref{Mathieu}. The form of the unperturbed solution, $\psi_0=\cos(2\theta)$ and $\beta_0=1$,
suggests to seek the perturbative solution in the form
\begin{subequations}
\begin{align}
\label{beta-epsilon}
\beta &= 1 + A \epsilon + \ldots\\
\label{psi-epsilon}
\psi &= \cos(2\theta) + \epsilon \psi_1(\theta)+ \ldots
\end{align}
\end{subequations}
Substituting \eqref{beta-epsilon}--\eqref{psi-epsilon} into \eqref{Mathieu} we obtain
\begin{equation}
\label{Mathieu-1}
\psi_1''+ \psi_1 +[8A+\cos(2\theta)]\cos(2\theta) = 0
\end{equation}
The symmetric solution to this equation reads
\begin{equation}
\psi_1 =  \tfrac{1}{12}\cos^2(2\theta)- \tfrac{1}{6} -2A\theta \sin(2\theta)
\end{equation}
where we omitted $C\cos(2\theta)$ term which can be absorbed into the unperturbed solution. The boundary condition $\psi_1(\pm \pi/4)=0$ fixes the amplitude $A=-1/(3\pi)$. Thus the decay exponent is
\begin{equation}
\beta = 1 - \frac{\epsilon}{3\pi} + O(\epsilon^2)
\end{equation}
Recalling \eqref{eps}, we thus have
\begin{equation}
\beta = 1 - \frac{2}{3\pi}\,\frac{Q}{D} +\ldots
\end{equation}

\subsection{Narrow wedges:  $\alpha<\alpha_*$}

In the general case of arbitrary opening angle in the range $\alpha\in (0,\alpha_*)$, the governing equation \eqref{phi-eq} with velocity field \eqref{u-Stokes} is again the Mathieu equation
\begin{equation}
\label{Mathieu-gen}
\psi''+ \{4\beta^2+\beta \epsilon [\cos(2\theta)-\cos(2\alpha)]\}\psi = 0
\end{equation}
with a slightly modified small parameter
\begin{equation}
\label{eps-alpha}
\epsilon = \frac{2Q}{D}\,\frac{1}{\sin(2\alpha) - 2\alpha\cos(2\alpha)}
\end{equation}
We seek the perturbative solution in the form
\begin{subequations}
\begin{align}
\label{beta-eps}
\beta &= \frac{\pi}{4\alpha} + A \epsilon + \ldots\\
\label{psi-eps}
\psi &= \psi_0 + \epsilon \psi_1+ \ldots, \quad \psi_0=\cos\!\left(\frac{\pi \theta}{2\alpha}\right)
\end{align}
\end{subequations}
The perturbation $ \psi_1(\theta)$ now obeys
\begin{equation}
\label{Mathieu-1-gen}
\psi_1''+ \left(\frac{\pi}{2\alpha}\right)^2 \psi_1 + \frac{\pi}{4\alpha}[8A + \cos(2\theta)- \cos(2\alpha)] \psi_0 = 0
\end{equation}
from which
\begin{eqnarray}
\psi_1 &=& \frac{\frac{\alpha}{\pi}\cos\!\left(\frac{\pi \theta}{2\alpha}\right)\cos(2\theta)+\sin\!\left(\frac{\pi \theta}{2\alpha}\right)\sin(\theta)\cos(\theta)}{(4\alpha/\pi)^2-4}\nonumber\\
&+&\theta\,\sin\!\left(\frac{\pi \theta}{2\alpha}\right)\left[\frac{1}{4}\,\cos(2\alpha)-2A\right]
\end{eqnarray}
The boundary condition $\psi_1(\pm \alpha)=0$ fixes the amplitude $A$. The decay exponent becomes
\begin{equation}
\beta = \frac{\pi}{4\alpha} + \frac{\epsilon}{16\alpha}\left[2\alpha\cos(2\alpha)+\frac{\sin(2\alpha)}{(2\alpha/\pi)^2-1}\right] + \ldots
\end{equation}
or equivalently
\begin{equation}
\beta = \frac{\pi}{4\alpha} + \frac{Q}{4D}\,\frac{\cos(2\alpha)+\frac{(2\alpha)^{-1}\sin(2\alpha)}{(2\alpha/\pi)^2-1}}{\sin(2\alpha) - 2\alpha\cos(2\alpha)} + \ldots
\end{equation}

 \subsection{Mean exit time}

The governing equation \eqref{tau-viscous} for the mean exit time becomes
\begin{equation}
\label{tau-Stokes}
\tau''+ \{4+\epsilon [\cos(2\theta)-\cos(2\alpha)]\}\tau = -1
\end{equation}
with $\epsilon$ given by \eqref{eps-alpha}. Plugging the perturbative solution
\begin{equation}
\tau = \tau_0+\epsilon\tau_1+\ldots
\end{equation}
into \eqref{tau-Stokes} we find
\begin{subequations}
\begin{align}
&\tau_0''+4\tau_0 = -1\\
&\tau_1''+4\tau_1 = - [\cos(2\theta)-\cos(2\alpha)]\tau_0
\end{align}
\end{subequations}
The boundary conditions are
\begin{equation}
\tau_0(\pm \alpha)=0, \qquad \tau_1(\pm \alpha)=0
\end{equation}
Solving for $\tau_0$ and then for $\tau_1$ we obtain
\begin{equation}
\label{tau-01}
\begin{split}
\tau_0 &= \frac{\cos(2\theta)-\cos(2\alpha)}{4\cos(2\alpha)}\\
\tau_1 &= \frac{\cos(4\theta)-3\cos(4\alpha)-6}{96\cos(2\alpha)} + \frac{1}{8}\,\theta\sin(2\theta) \\
&+\frac{3+\cos(4\alpha)-3\alpha\sin(4\alpha)}{48[\cos(2\alpha)]^2}\,\cos(2\theta)
\end{split}
\end{equation}
Thus in the Stokes regime a physically sensible solution for the mean exit time is only possible for acute opening angles $2\alpha<\frac{\pi}{2}$. The
mean exit time in this situation reads
\begin{equation}
\label{T:Stokes}
T(r,\theta)=\frac{r^2}{D}\,[\tau_0(\theta)+\epsilon\tau_1(\theta)+\ldots]
\end{equation}
with $\tau_0, \tau_1$ given by \eqref{tau-01} and $\epsilon$ given by \eqref{eps-alpha}. Interestingly, there are a few exact viscous similarity solutions in the wedge that also exist only for acute opening angles \cite{Fraenkel61,Moffatt80}.

\subsection{Wide wedges: $\alpha\geq \alpha_*$}
\label{subsec:wide}

The singularity of the velocity at $\alpha=\alpha_*$ has been noted by Fraenkel \cite{Fraenkel62}. The same angular dependence as in Eq.~\eqref{u-Stokes} has appeared in several  problems concerning wedges, e.g., in a problem \cite{Koiter58} related to elastic wedges and in fluid dynamics problems \cite{Moffatt64,Moffatt80}. A possible resolution of the singularity at $\alpha=\alpha_*$ relies on a more physical realization of the Jeffery-Hamel flow. Indeed, we have assumed that the size of the input region is equal to zero. We have used the boundary condition
\begin{equation}
\label{v:BC}
v|_{\theta=\pm \alpha}=0
\end{equation}
for the radial velocity and the boundary condition
\begin{equation}
\label{w:BC}
w|_{\theta=\pm \alpha}=0
\end{equation}
for the tangential velocity. To ensure that the flux has a non-zero strength $Q$ and emerges from an apex one must rely on generalized functions (distributions). Physically, the size of the input region is finite. Intriguingly, relying on this realistic property, one can overcome the singularity of the velocity at $\alpha=\alpha_*$.  To model the finiteness of the input region, it proves convenient \cite{Moffatt80} to replace the boundary condition \eqref{w:BC} by
\begin{equation}
\label{w:BC-a}
w(r,\theta=\pm \alpha)=
\begin{cases}
\mp \omega r   &0<r<a\\
0                      & r>a
\end{cases}
\end{equation}
The flux is thus introduced through the boundaries in the input region $r<a$ near the apex. The total flux is
\begin{equation}
Q = 2\int_0^a dr\, \omega r = \omega a^2
\end{equation}
The Jeffery-Hamel problem is recovered when the flux is introduced in the tiny region, more precisely in the limit $a\to 0$ and $\omega\to\infty$ with $Q=\omega a^2$ constant.

If $\alpha<\alpha_*$, the flow field far away from the input region ($r\gg a$) is the Jeffery-Hamel solution \eqref{u-Stokes} in the leading order. The stream function defining the radial and tangential velocity components via $v=r^{-1}\partial_\theta \psi, ~w=-\partial_r \psi$ is given by
\begin{equation}
\psi = \tfrac{1}{2}Q f(\theta), \quad f = \frac{\sin(2\theta)- 2\theta\,\cos(2\alpha)}{\sin(2\alpha) - 2\alpha\cos(2\alpha)}
\end{equation}
in the leading order. The details of the input region, viz. the parameters $a$ and $\omega$, do not affect the behavior.

\begin{figure}[ht]
\begin{center}
\includegraphics[width=0.4\textwidth]{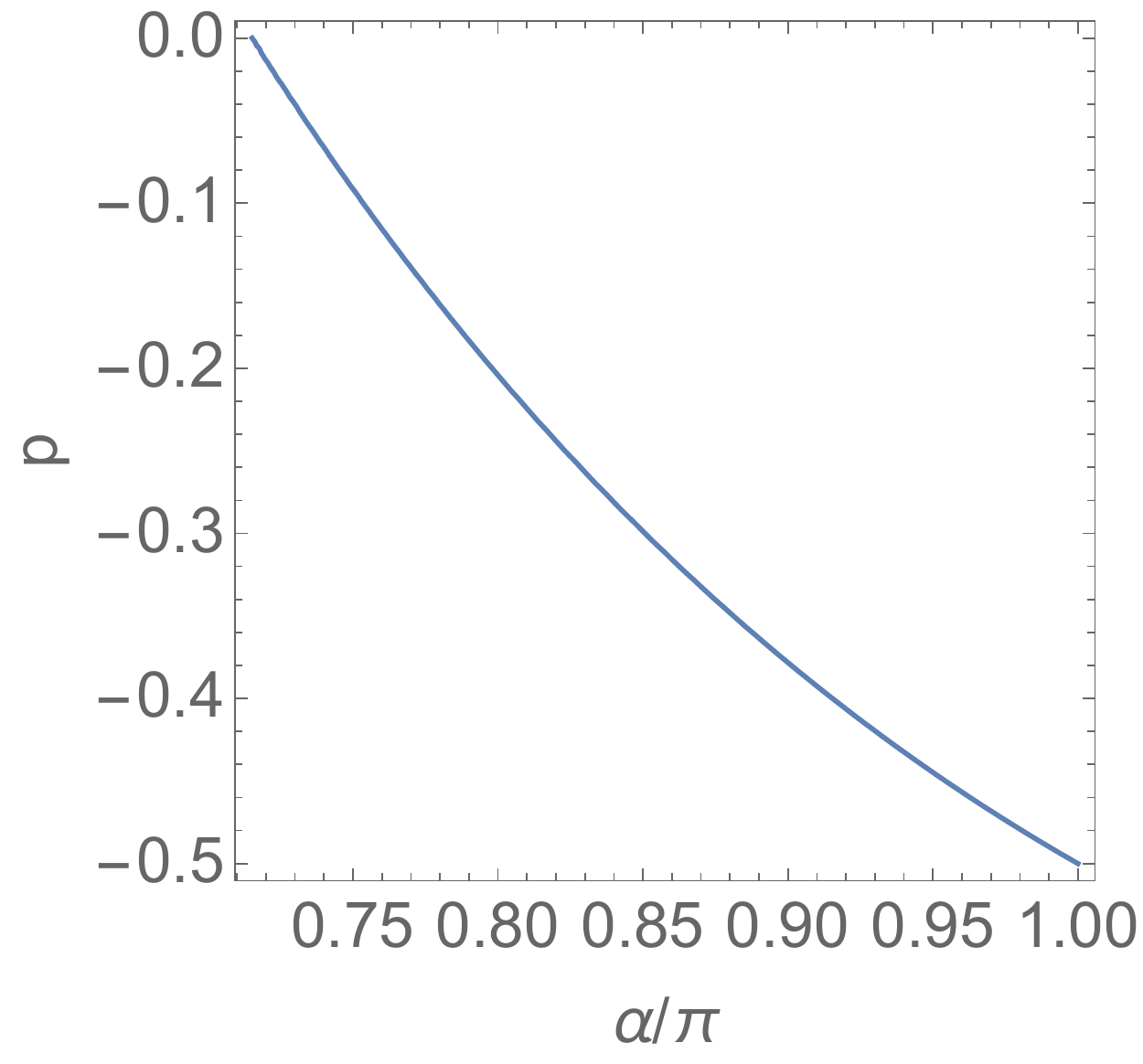}
\caption{The root $p=p(\alpha)$ of Eq.~\eqref{Wp-alpha} that determines the leading behavior \eqref{psi} of the flow field far away from the input region, $r\gg a$, for sufficiently wide wedges ($\alpha>\alpha_*$).}
\label{fig:p-alpha}
  \end{center}
\end{figure}

For $\alpha>\alpha_*$, however, the leading term is different, it depends on the details of the input region even far away from it. The stream function is \cite{Moffatt80}
\begin{equation}
\label{psi}
\psi = Q(r/a)^{-p}F(\theta)
\end{equation}
in the leading order. Here $p=p(\alpha)$ is the proper root of
\begin{equation}
\label{Wp-alpha}
W(p) \equiv (p+1)\sin(2\alpha)-\sin[2(p+1)\alpha] = 0
\end{equation}
There are many roots of $W(p)=0$, real and complex \cite{Moffatt80}; the proper root, $p=p(\alpha)$, is a decreasing function of angle in the $\alpha_*<\alpha<\pi$ range, with $p(\alpha_*)=0$ and $p(\pi)=-\frac{1}{2}$. The dependence of the stream function on the angular coordinate is \cite{Moffatt80}
\begin{equation}
\label{Fp}
F(\theta) = \frac{\cos[(p+2)\alpha]\,\frac{\sin p\theta}{p}- \cos[p\alpha]\,\frac{\sin(p+2)\theta}{p+2}}{W'(p)}
\end{equation}

The details of the hydrodynamic solution, such as the precise form of the angular dependence, Eq.~\eqref{Fp}, do not affect the ultimate fate of the diffusing particle. The only relevant features are the sign of the input $Q$ and the negativity of $p(\alpha)$ when in the $\alpha_*<\alpha<\pi$ range. For source flows, $Q>0$, the radial displacement grows as $t^{1/(p+2)}$, while the tangential scales diffusively as $t^{1/2}$. Since $p<0$, the radial displacement dominates, $t^{1/(p+2)}\gg t^{1/2}$. Thus we effectively have a diffusing particle in a growing interval with absorbing walls receding faster than diffusively. In this situation, the diffusing particle survives with a finite probability \cite{KR96,KR99}. The computation of $S_\infty(r,\theta)$ is difficult, but the chief property, $S_\infty(r,\theta)>0$ if $Q>0$, is clear. In the case of sink flows, $Q<0$, the particle is hovering on distances $r\sim R$ with
\begin{equation}
R\sim a\left(\frac{|Q|}{D}\right)^{1/p}
\end{equation}
(We tacitly assume that $\frac{|Q|}{D}\ll 1$ which is consistent with the low Reynolds number limit, $\frac{|Q|}{\nu}\ll 1$, in the natural situations when transport coefficients are comparable, $\nu\sim D$.) The survival probability is then exponential in time, $S\sim e^{-Dt/R^2}$, that is
\begin{equation}
S \sim \exp\!\left[-\frac{Dt}{a^2}\left(\frac{Q^2}{D^2}\right)^{-1/p}\right]
\end{equation}

Finally, a word of caution regarding the low Reynolds number limit.
The effective local Reynolds number
\begin{equation}
\text{Re} \sim \frac{Q}{\nu}\,(r/a)^{-p}
\end{equation}
eventually becomes large since $-\frac{1}{2}<p<0$. Therefore when $\alpha>\alpha_*$ the low Reynolds number treatment of the Jeffery-Hamel flow is inconsistent far away from the apex even if the {\em source} Reynolds number $Q/\nu$ is small.

\section{Three-dimensional analogs of the wedge flows}
\label{sec:3d}

An obvious three-dimensional analog is a flow in a circular cone. Unfortunately, the Navier-Stokes equations do not admit an analytical solution for the flow inside a cone. Fortunately, the decay exponent $\beta$ is independent of the flow. Indeed, if the particle has survived for a long time, it is far away from the apex. On large distances, the velocity field is $v\sim Q/r^2$, so advection is negligible compared to diffusion and the decay exponent $\beta=\beta(\alpha)$ is the same as if there were no flow \cite{SR,BK10a}. This decay exponent is the smallest root of the Legendre function:
\begin{equation}
\label{Legendre}
P_{2\beta}(\cos \alpha)=0
\end{equation}
At $\beta=\beta_c=1$, the opening angle is $\alpha_c=\text{Arccos}(1/\sqrt{3})$.

The decay exponent determined by Eq.~\eqref{Legendre} gives the ultimate large-time asymptotic. At intermediate times, the hydrodynamic flow field cannot be ignored. In Appendix~\ref{ap:cone} we outline some properties of the flow field. We rely on the Stokes approximation valid far away from the apex: $r\gg Q/\nu$. The analysis in Appendix~\ref{ap:cone} reveals that the hydrodynamic solution changes depending on whether  $\alpha$ smaller, equal, or larger than $\alpha_*=\frac{2\pi}{3}$.

A more interesting three-dimensional analog of the flow in the wedge is the jet flow caused by the point source of force (rather than mass as in the Jeffery-Hamel flow). For the jet flow the velocity components are also inversely proportional to the distance from the origin
\begin{equation}
\label{Landau}
v_r = \frac{\nu}{r}\,u(\theta), \quad v_\theta = \frac{\nu}{r}\,v(\theta)
\end{equation}
Although the velocity field \eqref{Landau} is not unidirectional, the $r^{-1}$ dependence of the velocity components makes the hydrodynamic problem solvable \cite{Slezkin34,Landau44,Squire51}.

The angular dependence of the velocity components is
\begin{equation}
\label{uv}
u=2\left[\frac{A^2-1}{(A-\cos \theta)^2}-1\right], \quad v =- \frac{2\sin\theta}{A-\cos \theta}
\end{equation}
and the pressure is given by \cite{LL-FM,Drazin}
\begin{equation}
p = p_\infty -\frac{4\nu^2}{r^2}\, \frac{A\cos\theta-1}{(A-\cos \theta)^2}
\end{equation}
The parameter $A$ is related to the momentum of the jet
\begin{equation}
M = 8\pi\nu^2A\left\{2+\frac{8}{3(A^2-1)}-A\ln\frac{A+1}{A-1}\right\}
\end{equation}
The quantity $\nu^{-1}\sqrt{M}$ plays the role of the Reynolds number of the jet flow.

Overall, the jet flow is a much better analog of the Jeffery-Hamel flow than the flow in the cone. The dimensionless momentum of the jet (equivalently, the Reynolds number $\text{Re}=\nu^{-1}\sqrt{M}$, or the parameter $A$) affects the exponent $\beta$. The same happens in the wedge where the dimensionless strength of the source (the P\'{e}clet number) affects the decay exponent $\beta$.

The analysis of the Brownian particle advected by the jet flow \eqref{Landau}--\eqref{uv} differs only in details from the analysis in the Jeffery-Hamel case. Suppose we want to determine the probability that a particle remains in the $\frac{\pi}{2} > \theta$ half-space during the time interval $(0,t)$. Using the same ansatz \eqref{decay} we find that $\Phi(r,\theta)$ satisfies
\begin{equation*}
\label{Phi-Landau}
\frac{\partial^2 \Phi}{\partial r^2} + \frac{2+\sigma u}{r}\,\frac{\partial \Phi}{\partial r} + \frac{1}{r^2}\,\frac{\partial^2 \Phi}{\partial \theta^2} + \frac{\cot\theta + \sigma v}{r^2}\,\frac{\partial \Phi}{\partial \theta} = 0
\end{equation*}
Seeking the solution in the form \eqref{Phi-sep} we arrive at
\begin{equation}
\label{phi-Landau}
\psi''+ (\cot\theta+\sigma v)\psi' +  2\beta(2\beta+1+\sigma u)\psi = 0
\end{equation}
Setting $\psi(\pi/2)=0$ ensures that the particle remains in the $\frac{\pi}{2} > \theta$ half-space.

The Sturm-Liouville equation \eqref{phi-Landau} can be reduced to the Schr\"{o}dinger equation. The physical requirement of positivity, $\psi>0$ when $0\leq \theta<\frac{\pi}{2}$, implies that we are seeking the ground state of a quantum particle in an infinitely deep potential well. The determination of the exponent $\beta$ may be easier than in the case of the wedge since the potential is expressible through trigonometric functions rather than elliptic functions. Leaving this to future work, we only consider weak jets. In this limit $A\gg 1$, so \eqref{uv} simplifies to
\begin{equation}
\label{uv-weak}
u=\frac{4}{A}\,\cos\theta, \quad v =- \frac{2}{A}\,\sin\theta
\end{equation}
and \eqref{phi-Landau} becomes
\begin{equation}
\label{phi-Landau-weak}
\psi''+ (\cot\theta-\delta\sin\theta)\psi' +  2\beta(2\beta+1+2\delta\cos\theta)\psi = 0
\end{equation}
with $\delta=2\sigma/A\ll 1$. When $\delta=0$ (no flow), $\beta_0=\frac{1}{2}$ and $\psi_0=\cos\theta$. For small $\delta$, we seek a perturbative solution
\begin{equation}
\label{beta-psi-jet}
2\beta = 1 + B \delta, \qquad \psi(\theta) = \cos \theta + \delta \psi_1(\theta)
\end{equation}
and find
\begin{equation}
\label{psi-1}
\psi_1''+ \cot\theta \psi_1' + 2\psi_1+1+(\cos\theta)^2+3B\cos\theta=0
\end{equation}
Solving this equation subject to $\psi_1(\pi/2)=0$ yields
\begin{equation}
B = -\frac{3}{4}\,, \qquad \psi_1 = \frac{(\cos\theta)^2-3\ln(1+\cos\theta)}{4}
\end{equation}
The general solution of Eq.~\eqref{psi-1} contains an extra term $(3+4B)\ln(1-\cos\theta)$, so the choice $B = -\frac{3}{4}$ ensures that $\psi_1$ remains regular on the axis of the jet ($\theta=0$). Thus
\begin{equation}
\beta=\frac{1}{2}-\frac{3\sigma}{4}\,A^{-1}+O(A^{-2})
\end{equation}
for weak jets.

\section{Conclusion}
\label{sec:concl}

We studied the first-passage characteristics of a particle diffusing in a wedge with absorbing boundaries, and advected by the flow generated by a source at the apex of the wedge. The survival probability decays algebraically with time. The decay exponent is easy to compute in the case of an ideal incompressible fluid. For the viscous fluid, we have reduced the determination of the exponent to finding the ground state energy of the quantum particle in an infinitely deep potential well. The shape of the well is determined by an exact solution of the Navier-Stokes equations for the incompressible viscous flow inside the wedge.

We employed perturbation techniques and deduced analytical predictions for the exponent $\beta$ describing the decay of the survival probability and for the mean exit time which is finite when $\beta>1$. The calculation of the mean exit time in planar domains is an active research subject \cite{FP,Schuss15,exit15,exit16,exit20a,exit20b,exit21}. Due to the scale invariance of the wedge, the dependence of $T(r,\theta)$ on the distance $r$ from the apex is fixed by dimensional arguments, Eq.~\eqref{T-sep}. The angular dependence is simple for ideal flows, Eq.~\eqref{T:ideal}. In the viscous case, we have established the angular dependence in the Stokes limit, Eqs.~\eqref{tau-01}--\eqref{T:Stokes}.

Amongst three-dimensional analogs of advection in the wedge, the closest is the jet flow (Sec.~\ref{sec:3d}). The analysis of the first passage characteristics is parallel to the analysis in the case of the wedge.

\bigskip\noindent
I am grateful to Alexei Skvortsov for correspondence.

\appendix
\section{Circular Cones}
\label{ap:cone}

In contrast to the wedge case, one cannot define a global Reynolds number in the case of the flow in a cone. Far away from the apex $v\sim Q/r^2$, so the local Reynolds number $rv/\nu\sim (Q/\nu) r^{-1}$ asymptotically vanishes irrespectively on the strength of the source. Thus, at least sufficiently far from the apex, one can employ the low Reynolds number approximation. In this $r\gg Q/\nu$ region, the velocity is purely radial: ${\bf v}=(v,0,0)$ in the spherical $(r,\theta,\phi)$ coordinates. The continuity equation and axial symmetry fixes the radial dependence,
\begin{equation}
\label{cone}
v = r^{-2}\,u(\theta),
\end{equation}
of the velocity. The Stokes equations are
\begin{equation}
\label{vr-Stokes}
\frac{1}{\nu}\frac{\partial p}{\partial r}=
\frac{\partial^2 v}{\partial r^2} + \frac{2}{r}\,\frac{\partial v}{\partial r} -\frac{2v}{r^2}+
\frac{1}{r^2\sin\theta}\,\frac{\partial}{\partial \theta}\left(\sin\theta \,\frac{\partial v}{\partial \theta} \right)
\end{equation}
and \eqref{pv} as in the case of wedge (where $(r,\theta)$ were polar coordinates). Using the ansatz \eqref{cone} we recast \eqref{pv} into
\begin{equation}
\frac{\partial p}{\partial \theta}=\frac{2\nu}{r^3}\,\frac{d u}{d \theta}
\end{equation}
which is integrated to yield
\begin{equation}
\label{pF}
p = \frac{2\nu}{r^3}\,u(\theta) + F(r)
\end{equation}
Substituting \eqref{cone} and \eqref{pF} into \eqref{vr-Stokes} we obtain
\begin{equation}
\label{Fu}
\nu^{-1}r^4\,\frac{dF}{dr} = u''+ u'\,\cot\theta+6u
\end{equation}
The left-hand side of \eqref{Fu} depends on $r$, while the right-hand side depends on $\theta$. Hence both sides must be equal to the same constant. In  particular
\begin{equation}
\label{u-eq-3}
u''+ u'\,\cot\theta+6u = \text{const}
\end{equation}
Solving \eqref{u-eq-3} subject to the no-slip boundary condition
\begin{equation}
u(\alpha)=0
\end{equation}
and the symmetry requirement
\begin{equation}
u'(0)=0
\end{equation}
one finds $u(\theta)=C[\cos 2\theta - \cos 2\alpha]$. The amplitude $C$ is fixed by mass conservation
\begin{equation}
Q = 2\pi \int_0^\alpha d\theta\,\sin\theta\,u(\theta)
\end{equation}
The final solution reads
\begin{equation}
\label{u-cone}
u(\theta)=\frac{3Q}{16\pi}\,\frac{\cos 2\theta - \cos 2\alpha}{(1+2\cos\alpha)\sin^4\frac{\alpha}{2}}
\end{equation}
Note that it becomes singular at $\alpha_*=2\pi/3$, so the solution \eqref{u-cone} is applicable only when $\alpha<2\pi/3$.

For wide cones, $\alpha\geq 2\pi/3$, the leading behavior is different from \eqref{cone} and \eqref{u-cone}. As in the case of the wedge (see Sec.~\ref{subsec:wide}) one can take into account the finiteness of the input region.


%

\end{document}